\pdfoutput=1
\documentclass[3p,times,twocolumn]{elsarticle}

\usepackage{ecrc}

\volume{00}

\firstpage{1}

\journalname{Nuclear Physics B Proceedings Supplement}

\runauth{}

\jid{nuphbp}

\jnltitlelogo{Nuclear Physics B Proceedings Supplement}

\usepackage{amssymb}
\usepackage[figuresright]{rotating}

\begin{document}

\begin{frontmatter}

 \title{Supernova Detection in IceCube: Status and Future}

\author{R. Bruijn for the IceCube Collaboration\fnref{www}}
\fntext[www]{http://icecube.wisc.edu}

\address{Laboratory for High Energy Physics, \'Ecole Polytechnique F\'ed\'erale, CH-1015, Lausanne, Switzerland}

\dochead{}

\begin{abstract}
The IceCube detector, located at the South Pole, is discussed as a detector for core collapse supernovae.
The large flux of $\bar{\nu}_{e}$ from a Galactic supernova gives rise to Cherenkov light from positrons and electrons created 
in neutrino interactions which increase the overall count rate of the photomultipliers significantly. We will 
give an overview of the standard, count rate based,
method for supernova detection and present the development of a novel technique. This technique uses coincident
hits to extract additional information such as the average energy and spectral features. The potential 
of this technique increases with a higher sensor density, such as foreseen in projected extensions of IceCube/DeepCore.

\end{abstract}

\begin{keyword}
Supernova \sep Neutrino \sep IceCube

\end{keyword}

\end{frontmatter}

\begin{figure}[ht]
\begin{center}
\includegraphics[width=0.9\columnwidth]{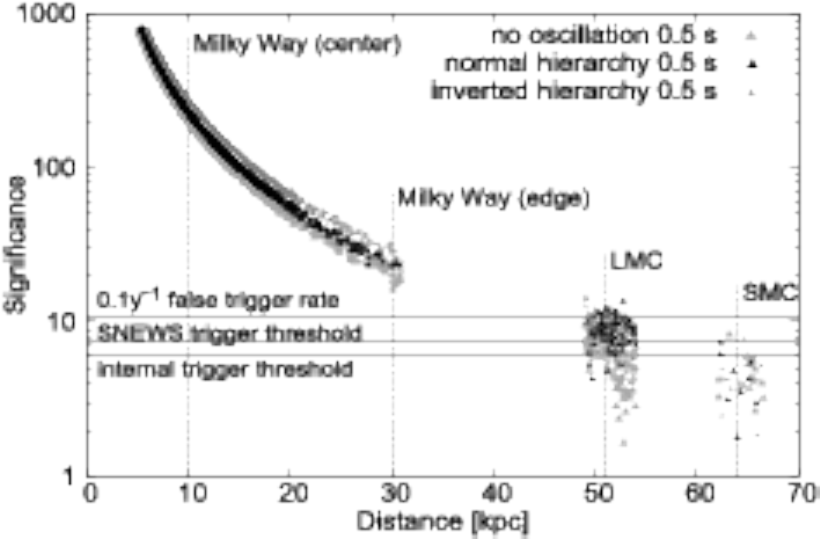}
\caption{\vspace{-1mm}Expected significance versus distance assuming the
Lawrence-Livermore model \cite{Totani:1997vj} for three oscillation scenarios. 
The density of the data points reflect the star distribution. (LMC/SMC: Large/Small Magellanic Cloud.)
 \vspace{-8mm}}
\label{fig:dist_reach}
\end{center}
\end{figure}

\begin{figure}[ht]
\begin{center}
\includegraphics[width=0.7\columnwidth]{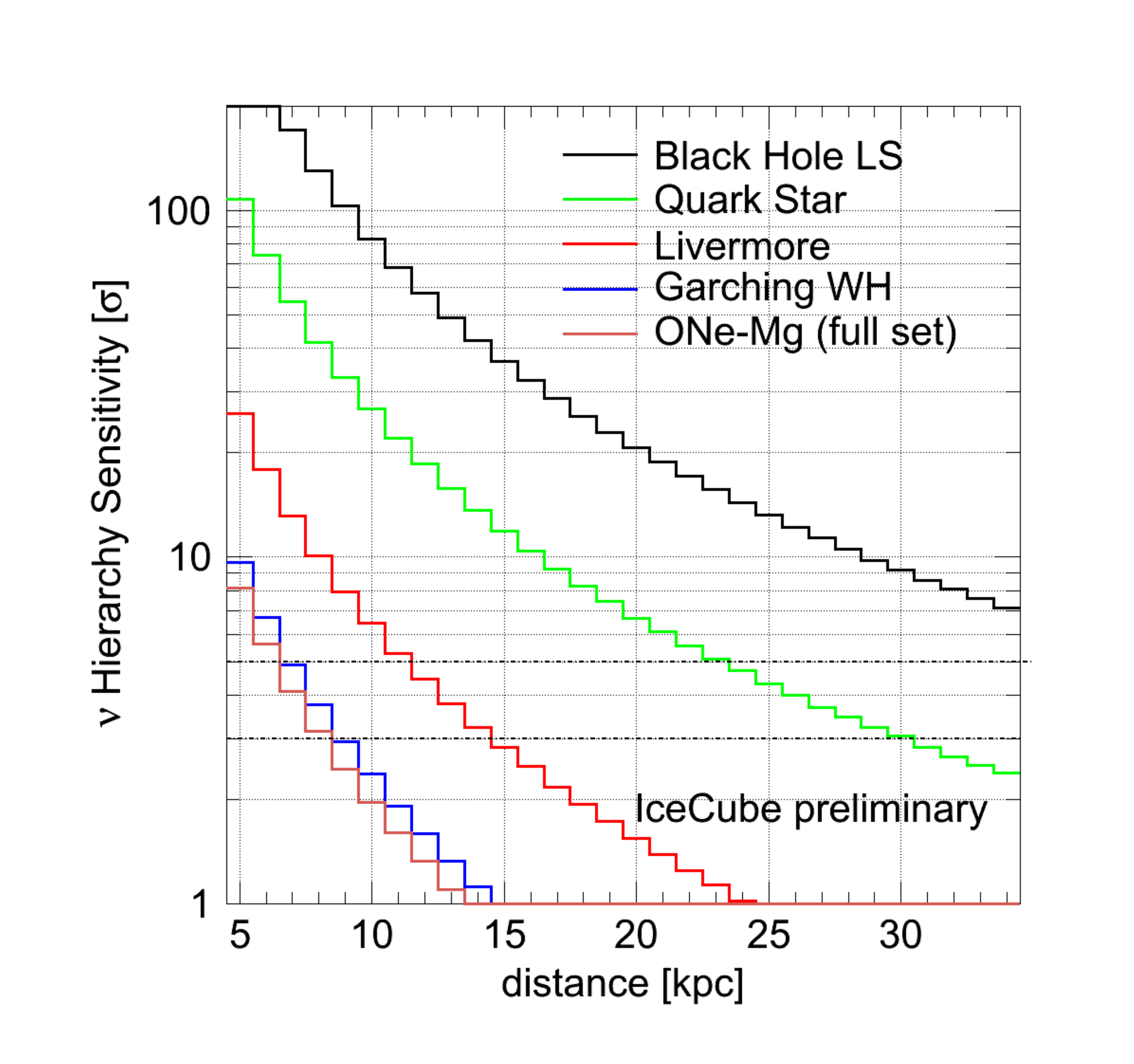}
\caption{\vspace{-1mm}Number of standard deviations with which normal
hierarchy and inverted hierarchy can be distinguished in at
least 50\% of all cases as function of supernova distance for several models.\vspace{-9mm}}
\label{fig:hierarchy}
\end{center}
\end{figure}

\begin{figure}[ht]
\vspace{-8mm}
\begin{center}
\includegraphics[width=0.7\columnwidth]{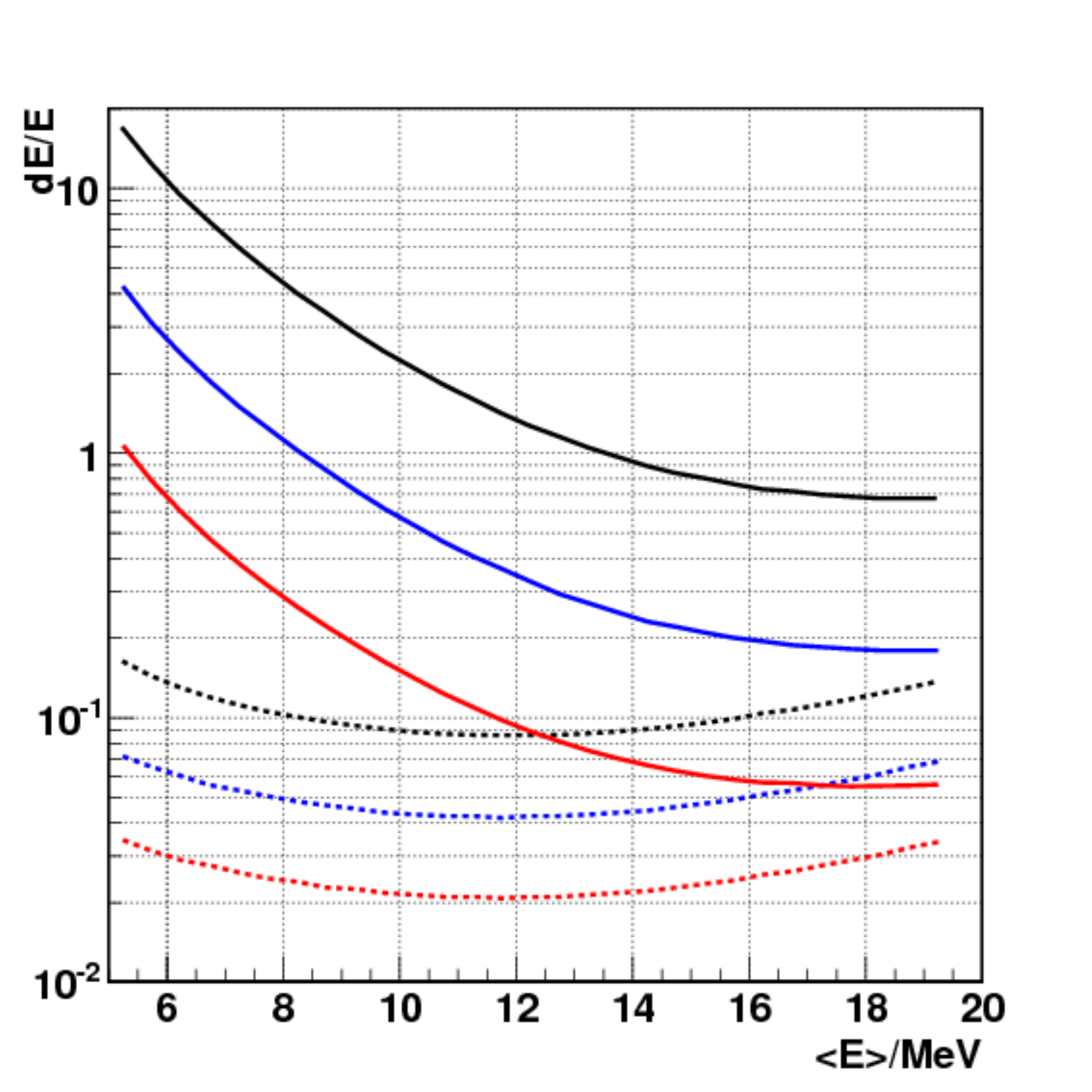}
\caption{\vspace{-1mm}Energy resolution, assuming a time integrated 8.8 solar mass O-Mg-Ne model with $\alpha=2.84$, for IceCube (solid) and
DDC (dashed) at a distance of  5 (red), 10 (blue)  and 20 (black) kpc.\vspace{-9mm}}
\label{fig:e_res}
\end{center}
\end{figure}

\begin{figure}[ht]
\vspace{-2mm}
\begin{center}
\includegraphics[width=0.7\columnwidth]{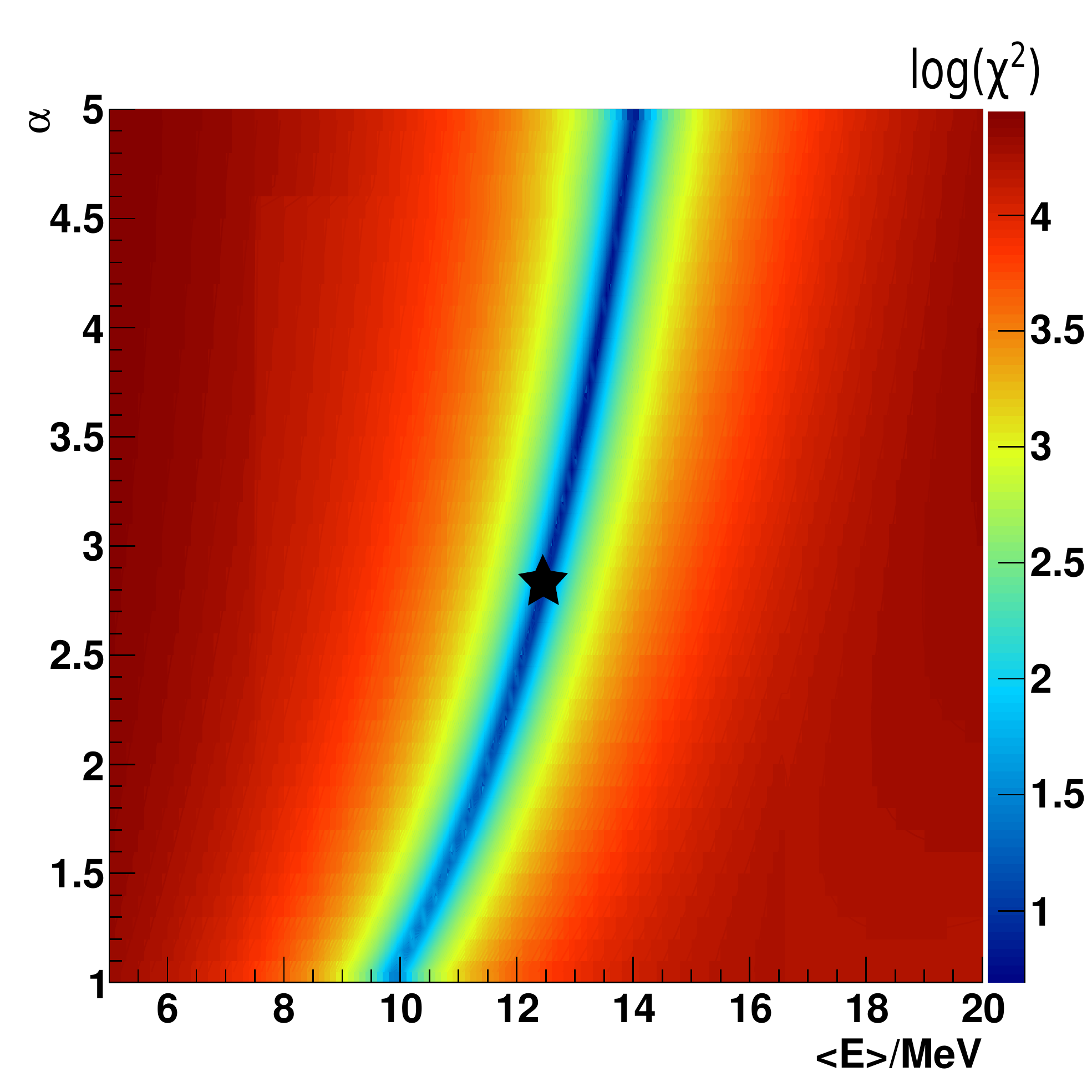}
\caption{\vspace{-1mm} $\log(\chi^{2})$ contour as function of energy and $\alpha$ for a 8.8 O-Mg-Ne supernova at 10 kpc and the DDC
detector geometry. True values indicated.\vspace{-11mm}}
\label{fig:e_a_scan}
\end{center}
\end{figure}

\emph{Introduction}\\
Many details of supernova explosions, like the exact mechanism, are still unknown, but it is clear 
that the largest
part of the gravitational energy is released in the form of neutrinos with an energy in the order of 10 MeV.
The detection of the neutrinos of a Galactic supernova can not only provide us with information
on the dynamics of the explosion \cite{Raffelt:2010zza}, but also on the nature of neutrinos such as 
the mass hierarchy.
While primarily designed to detect astrophysical TeV neutrinos, the 5160 10'' photomultipliers (PMTs) 
of the IceCube detector, instrumenting a cubic kilometer of ice,
are used to detect MeV neutrinos from a Galactic supernova \cite{Halzen:1994xe,Abbasi:2011ss}. The 
PMTs are housed 
in pressure-resistant glass spheres, together with read-out and digitization electronics, called
DOMs. The DOMs are organized in vertical strings, reaching from 1450 to 2450\,m depth. 
The 80 strings, each carrying 60 DOMs with a vertical spacing of 17.5\,m, are arranged in a hexagonal 
grid with a leg lengths of about 125 m.
Together with the surrounding strings, six additional strings carrying high quantum-efficiency PMTs, form the 
DeepCore sub-detector, which has a denser packing of about 60\,m between strings and 7\,m 
between DOMs on the same string.
The primary channel for the detection of supernova neutrinos is the inverse
beta-decay interaction of $\bar{\nu}_{e}$ on protons. The produced positrons, which travel about 5 cm at an energy of
10 MeV, emit approximately 2000 photons in the wavelength sensitivity range of the detector.
The reference supernova model has a 8.8 solar mass O-Mg-Ne progenitor \cite{Huedepohl:2009wh}  
and is considered conservative in terms of neutrino flux. 
This model is integrated over the first 3 seconds, optimizing for signal over background.
We will first give an overview and status of supernova detection in 
IceCube and continue with a relatively new method which uses coincident hits to extract additional 
information such as average energy and other spectral features.
This method is especially powerful with future projected dense detectors and results for a special
geometry, called \emph{Deep and Dense Core} (DDC) \cite{Demiroers:2011am}, will be presented. This geometry consists of 24 strings
with a 20\,m spacing each equiped with 120 4$\pi$ sensitive DOMs separated by 3\,m.\\
\\
\emph{Standard Method}\\
The detection of TeV neutrinos makes use of the long distance space/time correlations
between the photon hits caused by either a muon track or cascade initated by a neutrino.
The low energy of supernova neutrinos in combination with the large separation
of the DOMs make it impossible to reconstruct individual interactions. Instead, the detection technique
relies on the vast amount of interactions in the instrumented volume, about $8\times 10^7$  for the reference model at 10 kpc, 
which causes a significant increase in the count rate when considering all DOMs.
Key to this is the low noise rate of the individual DOMs of about 540 Hz, which is reduced to about 286 Hz after an artificial 
deadtime is applied after each hit, reducing the signal by 13\%. The noise level is continuously monitored
by a dedicated data-acquisition. It records scaler data in bins of 1.6384 ms. The data are later rebinned and synchronized
 in 2 ms bins. Following deployment, the noises rates show an exponential decrease which later levels out. A seasonal periodic behaviour
with an  amplitude of about 6\% is seen in addition. 
The most likely collective rate deviation $\Delta\mu$ from the baseline rate and its standard deviation $\sigma_{\Delta \mu}$ in a
time interval $\Delta t$ are calculated in real-time with respect to two sliding windows of 300\,s before 
and after the interval. Several values of $\Delta t$ (0.5, 4 and 10\,s) are used to take into
account different models. Triggers are set at different levels of the significance 
$\xi = \Delta\mu/\sigma_{\Delta \mu}$. When $\xi$ exceeds 6, an internal trigger is provided; when it exceeds 7.65, 
an alert is broadcasted on the Supernova Early Warning System (SNEWS). The
significance values can be corrected for a correlation with the atmospheric muon rate.
Significant triggers are caused by supernovae up to the Magellanic clouds (Fig. \ref{fig:dist_reach}).
By considering the shape and amplitude of the lightcurves, assuming the models are known, different
oscillation scenarios can be distinguished. Figure \ref{fig:hierarchy} shows the sensitivity to distinguish
between normal and inverted hierarchy for different explosion models.
The highest sensitivity is for scenarios in which a very massive star collapses into a quark star or black hole.
In these scenarios, total enery and energies of the neutrinos are particularly large.\\
\\
\emph{Luminosity and energy}\\
A limitation of using only the singles rate comes from the entanglement of the luminosity
and the energy distribution of the neutrinos \cite{Ribordy:2012gk}.
The flux of neutrinos $\tilde{\Phi}_{\bar{\nu}_{e}}$  at the 
detector from a supernova at distance $d$, can be described by
\begin{equation}
\frac{d\tilde{\Phi}_{\bar{\nu}_{e}}}{dE_{\nu}} =\frac{1}{4\pi d^{2}} \int^{\tilde{t}}_{0} 
\frac{L(t)}{\langle E_{\nu} \rangle(t)}f_{\alpha(t), \langle E \rangle(t)} dt
\label{eq:flux}
\end{equation}
with $L(t)$ describing the luminosity and $f_{\alpha(t), \langle E \rangle(t)}(E_{\nu})$ 
the normalized energy distribution, which depends on the average energy $ \langle E_{\nu} \rangle(t)$ and
a 'pinch' factor $\alpha(t)$. For $f_{\alpha(t), \langle E \rangle(t)}(E_{\nu})$, a commonly used parameterization
is utilized:
\begin{equation}
f_{\alpha(t), \langle E \rangle(t)}(E_{\nu})=\frac{(1+\alpha)^{(1+\alpha)}}{\langle E_{\nu} \rangle \Gamma(1+\alpha)}\left(\frac{E_{\nu}}{\langle E_{\nu} \rangle}\right)^{\alpha}e^{-(1+\alpha)\frac{E_{\nu}}{\langle E_{\nu} \rangle}}
\label{eq:f}
\end{equation}
The theoretical models provide  $L(t)$, $\langle E_{\nu}\rangle(t)$ and $f_{\alpha(t), \langle E \rangle(t)}(E_{\nu})$ 
for each neutrino species. For a measured supernova, these are unknowns
which have to be determined. In this work, we will assume that this formula can describe the
neutrino energy distributions. \\
From Eq.\ref{eq:flux}, it is evident that an increase in luminosity can be compensated by
a decrease in average energy. Furthermore,
the detection probability depends on the neutrino energy; the cross-section and light yield
increase quadratically and linearly with energy, respectively.
To disentangle the luminosity, energy and other spectral features, additional
energy sensitive observables are required.\\
\\
\emph{Coincident Hits}\\
The probability of photons from a single interaction to reach different DOMs increases with the 
energy of the neutrino. 
We consider coincidences of 1 (single rate) 2 or 3 DOMs, which can be nearest-neighbour,
 next-to-nearest-neighbour and on different strings. For 3-fold coincidences 1, 2 or 3
strings can be involved. The different hit modes are sensitive to different parts of the 
energy spectrum. This sensitivity is determined by the number of DOMs and their geometry. The
dependence of the probability is such that each extra DOM adds a factor of energy.
This sensitivity of coincident hits to the neutrino energy has been 
recognized in \cite{Demiroers:2011am} and is elaborated on in this work by including hit modes
with more than two DOMs and a full Monte-Carlo simulation based on Geant4 with custom
photon tracking. The atmospheric muon background is taken into
account, using a Gaisser parameterization of the flux.
The uncorrelated background is greatly reduced by the use of a small coincidence gate of 150 ns for IceCube,
100 ns for DeepCore and 50 for DDC. The correlated background from atmospheric muons can be reduced
by a multiplicity cut. The required raw hit information is made available by an improved
DAQ system which is currently being commissioned. Every hit is stored in a rotating 
filesystem, the \emph{hitspooling} system. In case of a supernova alert,
these data are transferred to permanent storage and are available for a 
detailed analysis, such as the coincident hit method.\\
\\
\emph{Determining energy and shape}\\
The energy dependent observables used are the ratios of the
rates of the coincidence modes to the rate of single hits, i.e. the singles rate is used
as an overall normalization, reducing systematic uncertainties. The ratio of the two-fold nearest-neighbour rate
to the singles rate is indicated by $r^{11}_{10}$. Assuming the functional form of the energy distribution in Eq.\ref{eq:f} with 
a fixed value $\alpha$, the energy dependence 
of $r^{11}_{10}$ on the average neutrino energy can be used to determine this energy from the rate \cite{ribordy_icrc_2009}.
The energy resolution  obtained for IceCube and DDC using $r^{11}_{10}$, can be seen in Fig. \ref{fig:e_res}. 
With IceCube, a resolution of 30\,\% can be obtained for a supernova at 10 kpc. The power of the method,
when used with a dense detector, is clear. With DDC, the resolution for the same conditions
is improved to 4\,\%. By combining ratios, information on the
spectral shape can be obtained. A first approach is constructing a $\chi^{2}$ test which includes
ratios of all mentioned rates, and takes into account the correlations between the modes. This $\chi^{2}$ 
quantifies the difference between a measurement (simulation, in this work)
and a chosen model (set of parameters $\alpha$ and $\langle E \rangle$).
A scan over the $(\langle E \rangle,\alpha)$ parameter space, presented in fig. \ref{fig:e_a_scan}, shows that
the parameters can be constrained, however there is a degeneracy between them.\\



\end{document}